\documentclass{article}
\usepackage{iclr2027_conference,times}
\iclrfinalcopy   
\usepackage{booktabs}
\usepackage{graphicx}
\usepackage[normalem]{ulem}
\usepackage{hyperref}
\hypersetup{hidelinks}
\usepackage{url}

\title{Audit Without Verification: When LLM Accountability Layers Relay Rather Than Check}

\author{Paul-Peter Arslan \\ Institute For Future Technologies \\ \texttt{paulpeterarslan@gmail.com}}

\begin{document}
\maketitle
\lhead{Preprint. Under review.}

\begin{abstract}
Multi-agent LLM pipelines increasingly span organisational boundaries, and when a fault surfaces someone must determine where it entered. In deployment the artifact available for that determination is rarely a full execution trace: it is the reports each agent filed, and a filed report can state a conclusion alongside its observations. We ask what an accountability layer built on such reports can and cannot do. Using a pre-registered, institutionally partitioned pipeline of six agents with process-level information boundaries, exactly balanced defect injection and matched clean twins (345,600 requests per chain model, two models), we first report that our pre-registered hypothesis --- that collective responsibility framing degrades escalation increasingly with chain length --- is not supported.

The layer nevertheless fails, and it fails \textbf{asymmetrically}. It originates almost nothing: zero allegations across 7,996 clean episodes where every agent stayed silent. It filters upstream error poorly, naming an innocent party in 34.4\% and 62.6\% of clean episodes where an agent raised a false alarm. An auditor reading agents' reports names an origin from within the set the agents proposed in 95\% of episodes; \textbf{conditional on no agent proposing the true origin} (59.5\% of episodes on one chain model) it recovers that origin in 4.1\% of cases --- below both a uniform guess (20\%) and the best fixed-link accuser in that stratum (31.0\%) --- while reaching 60.3\% from the raw documentation of the same episodes.

Deleting one clause, the field carrying the agents' own conclusion, isolates the cause at constant observations. Within that stratum accuracy rises to 45.2\% (+41.2 pp, 95\% CI +35.3 to +46.9) and adherence collapses from 94.4\% to 3.4\%; where the agents' suggestion was correct the same deletion instead costs accuracy, 70.5\% to 55.7\%. Because the two strata pull in opposite directions, the unconditional effect is small by construction, and we report the conditional quantities throughout. The effect replicates on a second auditor and, in a check pre-specified before its collection, on two frontier auditors in four conditions out of four (+8.5 to +39.0 pp, all intervals excluding zero). In a \textbf{second domain} --- a multi-vendor software delivery toolchain, replicated under a protocol frozen before collection --- the harm is larger still (+47.7 and +61.1 pp) while the cost disappears, both correct-suggestion intervals including zero: the trade-off is domain-dependent.

The finding is therefore not that removing the field is better. It is that the net effect is governed by upstream reliability \textbf{together with} the conditional benefit of suppressing an incorrect conclusion and the conditional cost of suppressing a correct one --- $\Delta$ = P(wrong)$\cdot$$\Delta$$\mid$wrong + P(correct)$\cdot$$\Delta$$\mid$correct --- and that upstream reliability can be estimated offline on representative validation data. An accountability layer needs evidence sufficiently independent of the conclusions it verifies.

\end{abstract}

\section{Introduction}
Agent pipelines are moving from single-vendor demonstrations to deployments in which distinct organisations own distinct steps: one company's model prepares an order, another manufactures against it, another inspects, another ships. Each controls what its agent sees and what it passes on. When a fault reaches the end of such a chain, the accountability question is not `\texttt{did the system fail'' but }`at which step, under whose control, did the fault enter''. This paper measures one concrete obstacle to answering it.

Automated failure attribution has a benchmark ecosystem, and it has established that \textit{how much} of the execution record an attributor sees matters: full traces substantially outperform output-only ones. We take that as settled and study a different property of the record. A filed report is not a truncated trace: it is written to be acted on, and can state a conclusion --- this step looks wrong --- alongside the observations supporting it. We hold the record fixed and vary whether it carries that conclusion. We make no claim that other benchmarks' traces lack such conclusions; the distinction is that varying this field as a controlled condition, at constant observations, has not been done, and the situation arises precisely where a pipeline crosses organisational boundaries and no shared trace exists.

Our claim is not that language models are susceptible to anchoring. That is established, and the mechanism here is consistent with it. What a bias account does not supply is what we report: that the layer is \textbf{asymmetric}, almost never originating an allegation yet endorsing a large share of upstream false ones, so that it behaves as a relay rather than a check; that the dependence has a \textbf{single removable carrier} in the report schema, identified by deleting one clause at constant observations; and that removing it is \textbf{not an improvement but a trade-off}, whose net effect depends on upstream reliability and on the conditional benefit and cost of suppression. A bias framing implies the field should go. Our measurements say when.

Varying the number of principals isolates no mechanism: principal count is entangled by construction with organisational boundaries, segment length, and the share of steps under common control, and adjusting for those adjusts for consequences of the manipulation, not confounds of it. We therefore vary the proposed mechanism --- how responsibility for flagging anomalies is allocated --- factorially against principal count and against a semantically matched control domain, holding monitoring scope, blame attribution and the operational reporting duty textually constant across every cell.

\textbf{RQ1 (pre-registered).} Does the individual-versus-collective escalation gap widen as chain length grows, specifically for responsibility framing rather than for matched language about documentation?

\textbf{RQ2.} How much information about a fault's origin survives the reporting layer, and is the loss caused by the report format or by the auditor's own limitations?

RQ1 is the registered confirmatory test and is answered in the negative. RQ2 is where the result lies, and its provenance is mixed in a way we state rather than smooth over. The auditor analysis and the restricted-versus-full contrast were specified before collection as explicitly descriptive; the restricted interface deviates from the registered implementation (§3.4), so we never label that comparison simply ``pre-specified''. The single-clause manipulation is an exploratory follow-up and the least affected by that deviation, comparing one report format against the same format minus one clause --- which is why the causal weight rests there.

\textbf{Contributions.} (1) A pre-registered three-factor design with blocked randomisation, matched clean twins and the discrimination index J, reported in full including its null primary result. (2) A characterisation of the accountability layer as \textbf{asymmetric}: it rarely originates an allegation absent upstream escalation, yet accepts a substantial fraction of erroneous ones. (3) A single-clause causal manipulation identifying the conclusion-bearing field as a \textbf{removable carrier} of that dependence at constant observations, with the residual gap to raw documentation reported rather than absorbed into the claim, replicated on a second auditor and on two frontier auditors. (4) The trade-off characterised \textbf{in both directions}, its net effect decomposed into upstream reliability and the two conditional magnitudes, with an offline instrument for estimating it. (5) An information-theoretic attribution metric read against a constant-accuser baseline exact at the design level, with within-stratum comparators reported where conditioning breaks that balance.

\section{Related work}
\textbf{Failure attribution.} \citet{zhang2025whowhen} established automated failure attribution as a task, releasing failure logs from 127 multi-agent systems annotated with the responsible agent and the decisive error step. Attribution difficulty is established there and is not our claim: their best method reaches 53.5\% agent-level but only 14.2\% step-level accuracy, with some methods falling below random. Observability is likewise a studied axis, and we do not claim otherwise: \citet{chen2026traceelephant} note that prior benchmarks expose only partially observable traces capturing agent outputs, introduce a full-execution-trace benchmark, and report gains of up to 76\% over a partial-observation counterpart. \textbf{How much of the record an attributor sees therefore already has a literature with a measured effect; our contribution is not to rediscover it.} What we vary instead is whether the record additionally carries an explicit upstream attribution, at constant observations.

We report mutual information alongside accuracy, which separates a noisy attributor from one emitting a near-fixed answer --- two systems that can post identical accuracy.

\textbf{Adoption of upstream claims.} That agents adopt peers' answers to their own cost is measured: \citet{hao2026conformity} decompose stance convergence in multi-agent debate and find strict conformity predominantly harmful, 57--77\% of it correct-to-wrong, with even vacuous reasoning driving 20--39\% error adoption among otherwise resistant agents. \citet{wang2026provenance} survey evidence tracing and execution provenance as foundations for process-level accountability in LLM agents.

Our contribution is not to observe that upstream claims are adopted, nor that provenance matters. It is to identify a specific, removable carrier of the dependence in a deployed accountability interface: a literature establishing that peers' assertions are adopted does not tell a designer which line of their report schema is transmitting one.

\textbf{Anchoring.} Anchoring in LLMs is a known phenomenon with dedicated work on its existence, mechanism and mitigation: \citet{huang2025anchoring} report that it arises commonly, acts at shallow layers, and resists conventional mitigation strategies. The mechanism underlying our result is consistent with this literature, and a reader who concludes ``this is anchoring'' is not wrong about it. Our claim concerns its locus and consequence: \textbf{where} it occurs (the layer a deployed system uses to assign responsibility across organisational boundaries, not a pairwise preference judgment); \textbf{how large} it is against a baseline fixed in advance (4.1\% against 20\%, below what naming a fixed link would score); \textbf{what causes it} (deleting one clause while leaving the observations intact recovers 41.2 points); and \textbf{what removing it costs} (14.7 points where the upstream suggestion was correct --- anchoring framed purely as bias implies removal is an improvement; measured as an interface trade-off it is not).

\textbf{Protocol security.} Security of tool-using agent protocols is an active area: \citet{huang2026mcpthreat} threat-model Model Context Protocol implementations and evaluate tool-poisoning defences across seven clients. This is \textbf{not} a protocol-security paper: we present no attack and no defence. We use per-company MCP servers because they enforce information boundaries at the process level rather than by instruction.

\section{Method}
\textbf{3.1 Environment.} A six-link supply-chain process (raw-material certification, manufacture, quality inspection, freight, distribution intake, retail intake). Each link is handled by one agent that completes its step and files a structured report. A defect is injected at exactly one link \textit{i} $\in$ \{1,\ldots,5\}; link 6 is never an injection site, since no downstream agent could observe it. Two defect families: \textbf{inconsistency} (a quantity contradicting figures already in the record) and \textbf{omission} (a required record absent). A third was generated and excluded before collection, both models having flagged it at 100\% per link in calibration.

Injection positions are exactly balanced by construction --- across 480 base cases each of positions 1--5 occurs 96 times --- and case content is a pure function of a recorded seed, so the corpus is regenerable without redistribution.

Human corpus validation on a blind stratified sample, against criteria fixed before any answer existed, gave \textbf{status accuracy 100\%} and \textbf{unintended defects 0\%}. A third criterion failed at 70\% against a 90\% gate; a follow-up traced this to a conflated question and found the stated fact unambiguous in 100\% of cases. A single rater was used (Appendix A).

\textbf{3.2 Architecture.} Each company identity runs as a separate MCP server process coordinated by an orchestrator client. What an agent can see is set by which tools are registered on its server, not by an instruction to disregard visible information; under the visibility condition used throughout, the topology tool is absent from the advertised list and the build asserts this. We describe the system as a \textbf{synthetic institutionally partitioned agent pipeline with enforced information boundaries}; this is not a claim about genuine multi-principal economics, and the study cannot distinguish sophisticated instruction-following from motivation.

Model output is never fed forward: the orchestrator grows the work product by the case's defect text and a fixed completion marker only, verified byte-for-byte against the live execution path. The agents' reports are therefore the \textit{only} channel by which one agent's interpretation could reach an auditor.

\textbf{3.3 The pre-registered factorial.} Three factors, twelve cells: principals \textit{k} $\in$ \{1,3,6\} $\times$ domain $\in$ \{responsibility, documentation\} $\times$ allocation $\in$ \{individual non-delegable, collective shared\}. One base case is replayed across all twelve cells and a matched no-defect twin, removing case difficulty from every within-block comparison: 480 blocks $\times$ 12 cells $\times$ 2 twins $\times$ 5 replications $\times$ 6 links = \textbf{345,600 requests per chain model}, giving 57,600 episodes. Held constant across every cell: monitoring scope, blame attribution, the operational reporting duty, chain length, role descriptions, output schema, temperature and case content. The allocation manipulation is one paragraph of framing text, in three lexical variants rotated across blocks; the documentation domain is the semantic control, holding collective-versus-individual language constant while removing responsibility content.

The primary outcome is \texttt{episode\_escalated}, a logical OR over the six links; an episode with any invalid link is dropped whole and counted. Matched twins make the false-alarm rate estimable within each cell, licensing J = P(escalate $\mid$ defective) - P(escalate $\mid$ clean), modelled inside one saturated GEE rather than by subtracting two fits.

\textbf{3.4 The accountability layer, and a declared deviation.} The pre-registration fixed, before any confirmatory data existed, that an external auditor would run in a \textbf{communications-only} and a \textbf{full-information} condition, the attribution scoring, the requirement that the auditor be a different model from the chain agents, and the 20\% naive baseline --- as explicitly descriptive analyses.

The restricted interface actually used \textbf{deviates from the registered implementation}, and we do not describe the comparison as simply `\texttt{pre-specified''. The registered text describes communications-only as receiving }\texttt{only the messages passed between links''. \textbf{No such messages exist in this architecture} (§3.2), so the registered description was not implementable as written and the departure was forced at implementation time rather than chosen after seeing results. We operationalised it as each link's own filed report --- escalation flag, anomaly type, stated suspected origin, free-text evidence --- plus the initial documentation common to every condition. One consequence, stated plainly rather than defended: this gives the auditor each agent's stated conclusion, which a literal }`messages between links'' condition might not have. Since our finding is that supplying such a conclusion degrades attribution, the deviation is not neutral with respect to §4.2 and may be part of what that comparison measures. This is why the causal weight rests on the single-clause manipulation.

A second registered analysis \textbf{was not conducted}: the evidence-content rubric required validation against two independent human raters, that validation was not performed, and we report no result from it. A model self-rating collected in the same call is reported nowhere: the same model rates the evidence and produces the attribution, so a failed attribution can itself produce an ``absent evidence'' rating. The measure is circular here.

\textbf{3.5 The single-clause manipulation (exploratory).} To isolate the conclusion-bearing field we delete exactly one clause from the communications-only report:

\begin{quote}
\texttt{ESCALATED. Anomaly type: <type>.} \sout{Suspected origin: step N.} \texttt{Report: <free text>}
\end{quote}
The struck clause is the only thing the manipulation removes. The escalation flag, anomaly type, free-text report, initial documentation, auditor model, schema, temperature and episode set are unchanged; the two prompt versions were diffed programmatically to confirm nothing else differs. Two controls in §4.3 answer the obvious alternatives: residual cues in the free text, and the direction of the abstention rate.

\textbf{3.6 Statistics.} GEE \citep{liang1986gee}, binomial/logit, clustered by block, with the Mancl-DeRouen bias-corrected sandwich \citep{mancl2001sandwich}; replications enter as individual Level-1 observations; each chain model is fit separately and never pooled; no mediator covariates; contrasts reported as marginal probability differences. \textbf{Deviation:} the registered exchangeable working correlation did not converge; we report an independence working correlation with the same block-clustered sandwich, valid under \citet{liang1986gee}, rather than substituting silently. Appendix G gives the contrast under four specifications, including a random-intercept GLMM; all include zero. Exploratory contrasts use paired block-clustered bootstraps over the 480 base cases.

\textbf{3.7 Provenance of each analysis.}

\begin{center}
\footnotesize
\setlength{\tabcolsep}{3.0pt}
\begin{tabular}{p{0.460\linewidth}p{0.490\linewidth}}
\toprule
\textbf{analysis} & \textbf{status} \\
\midrule
three-way contrast; J-contrast & \textbf{confirmatory} \\
per-cell detection / false alarm / J & pre-specified descriptive \\
auditor: restricted vs full information & pre-specified comparison; interface deviated \\
single-clause deletion; replication; clean-episode study & \textbf{exploratory} \\
allocation effect; mutual information & \textbf{exploratory} \\
frontier-auditor check and second-domain replication (§4.4) & \textbf{each pre-specified before its own collection}; neither in the original pre-registration \\
free-text leakage control; defect-family and position breakdowns & \textbf{exploratory}, after review \\
interleaved re-collection (§4.3) & \textbf{pre-specified control}, after review; same episodes, single execution \\
evidence-content rubric & registered; \textbf{not conducted} \\
\bottomrule
\end{tabular}
\end{center}

\textbf{3.8 Models.} Chain agents: \texttt{gpt-4.1-mini} and \texttt{claude-haiku-4-5}, fit separately as independent replications. Auditors: \texttt{qwen3-30b-a3b-instruct} (primary) and \texttt{gemini-2.5-flash-lite} (replication), plus \texttt{gpt-5} and \texttt{claude-sonnet-5} for the frontier check of §4.4; all distinct from both chain models. This reduces the risk of same-model coupling between acting and auditing roles; it does not eliminate correlated error, since independently developed models can share biases. A screen of seven additional low-cost models found five below a 5\% per-link escalation rate and three returning \texttt{escalate: false} on 300 of 300 calls, against references of 17.1\% and 12.5\% (Appendix C): the corpus difficulty was calibrated against the two models used and does not transfer downward.

\section{Results}
\subsection{The pre-registered contrast is null (confirmatory)}
\begin{center}
\footnotesize
\setlength{\tabcolsep}{3.0pt}
\begin{tabular}{p{0.283\linewidth}p{0.141\linewidth}p{0.318\linewidth}p{0.177\linewidth}}
\toprule
\textbf{chain model} & \textbf{contrast} & \textbf{95\% CI} & \textbf{excludes 0} \\
\midrule
GPT-4.1 Mini & +0.0108 & [-0.0305, +0.0522] & no \\
Claude Haiku 4.5 & -0.0129 & [-0.0411, +0.0152] & no \\
\bottomrule
\end{tabular}
\end{center}

The registered J-contrast agrees: GPT +0.0087 [-0.0387, +0.0562]; Claude -0.0183 [-0.0480, +0.0114]. We state this plainly as a null result; the diffusion-of-responsibility account is not invoked again as though this test had supported it. Reporting it first, rather than after the results that worked, is what makes the provenance labels in the rest of this section credible.

\subsection{Provenance degrades through the reporting layer, and the auditor adopts the chain's origin}
Same auditor, same episodes, same ground truth; only the input differs.

\begin{center}
\footnotesize
\setlength{\tabcolsep}{3.0pt}
\begin{tabular}{p{0.161\linewidth}p{0.211\linewidth}p{0.080\linewidth}p{0.181\linewidth}p{0.271\linewidth}}
\toprule
\textbf{chain} & \textbf{auditor reads} & \textbf{accuracy} & \textbf{mutual information} & \textbf{within agents' proposed set} \\
\midrule
GPT-4.1 Mini & agents' filed reports & 26.0\% & 0.065 bits & \textbf{95.0\%} \\
GPT-4.1 Mini & raw documentation & 63.9\% & 1.073 bits & 47.6\% \\
Claude Haiku 4.5 & agents' filed reports & 63.2\% & 0.809 bits & \textbf{95.0\%} \\
Claude Haiku 4.5 & raw documentation & 63.3\% & 1.057 bits & 65.3\% \\
\bottomrule
\end{tabular}
\end{center}

Maximum available information is $\approx$2.3 bits; the constant-accuser baseline is 20\% and 0 bits. Mutual information is exploratory. Adherence, used throughout: the auditor's named origin lies within the \textit{set} some agent proposed, both sides restricted to links 1--6. We attach no null agreement rate --- agents' proposals are not uniformly distributed (Appendix B).

\textbf{The decisive stratification (exploratory).} Partitioning by whether the true injection link appears among any agent's proposal; both strata reported.

Where the truth \textbf{is} proposed (4,003 and 6,468 episodes), the auditor recovers it from reports in 70.5\% and 80.6\% of cases, against 76.7\% and 69.1\% from raw documentation. Where it is \textbf{not} (5,875 and 2,414 episodes --- 59.5\% of GPT-chain episodes), the same auditor falls to \textbf{4.1\%} and \textbf{8.7\%} from reports while raw documentation still yields \textbf{60.3\%} and \textbf{69.4\%}.

\textbf{The exact 20\% balance does not survive this conditioning, and we do not claim it does.} The stratum is defined by model behaviour, not design, and that behaviour correlates with the true position: agents name link 1 constantly, so episodes whose true origin \textit{is} link 1 almost always have it proposed and are removed from this stratum. True positions within it are consequently skewed --- link 1 falls to 6.7\% and link 5 rises to 31.0\% (GPT chains; 6.8\% and 43.8\% for Claude). We therefore use within-stratum comparators rather than carrying the design-level baseline across. The auditor falls below both: \textbf{4.1\%} against a 20.0\% uniform guess over the five sites and a \textbf{31.0\%} best fixed-link accuser on GPT chains; \textbf{8.7\%} against 20.0\% and \textbf{43.8\%} on Claude chains --- the wider margin being against the stronger comparator. At n = 5,875 this is not a small-sample artifact; it indicates attraction toward supplied incorrect suggestions rather than independent noisy reasoning. The stratification is exploratory and defined by the agents' own outputs, not randomised.

\subsection{Removing one clause (exploratory --- the causal centre)}
\label{sec:strip}

\begin{figure}[t]
\centering
\includegraphics[width=\linewidth]{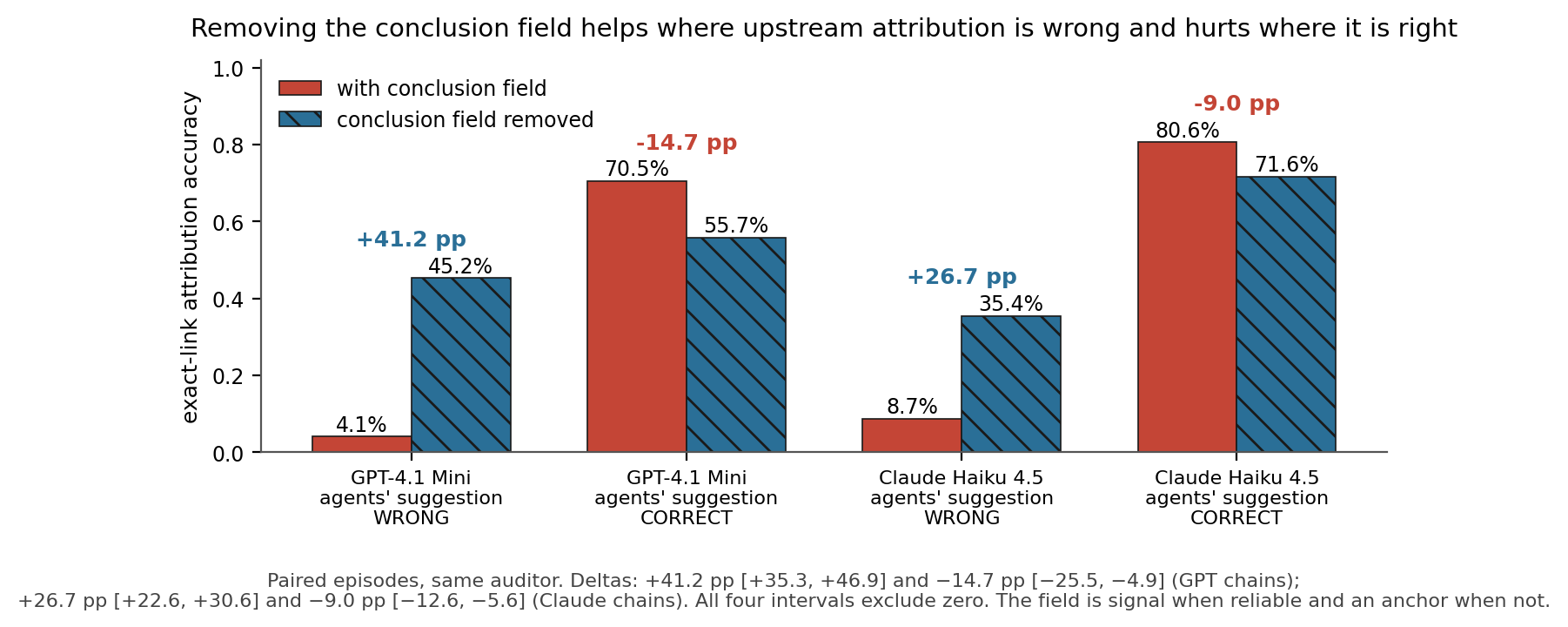}
\caption{The central result and its trade-off. Removing the conclusion field helps where upstream attribution is wrong and hurts where it is right; all four paired intervals exclude zero.}
\label{fig:tradeoff}
\end{figure}

The §4.2 comparison differs in many ways at once and cannot identify a cause. Deleting exactly one clause, and nothing else:

\begin{center}
\footnotesize
\setlength{\tabcolsep}{3.0pt}
\begin{tabular}{p{0.170\linewidth}p{0.191\linewidth}p{0.069\linewidth}p{0.106\linewidth}p{0.095\linewidth}p{0.095\linewidth}p{0.148\linewidth}}
\toprule
\textbf{chain} & \textbf{agents' suggestion} & \textbf{n} & \textbf{with field} & \textbf{without} & \textbf{$\Delta$} & \textbf{95\% CI} \\
\midrule
GPT-4.1 Mini & \textbf{wrong} & 5,864 & 4.1\% & \textbf{45.2\%} & \textbf{+41.2} & [+35.3, +46.9] \\
GPT-4.1 Mini & correct & 3,988 & 70.5\% & 55.7\% & \textbf{-14.7} & [-25.5, -4.9] \\
Claude Haiku 4.5 & \textbf{wrong} & 2,407 & 8.7\% & \textbf{35.4\%} & \textbf{+26.7} & [+22.6, +30.6] \\
Claude Haiku 4.5 & correct & 6,439 & 80.6\% & 71.6\% & \textbf{-9.0} & [-12.6, -5.6] \\
\bottomrule
\end{tabular}
\end{center}

All four intervals exclude zero: both directions of the trade-off are estimated, not merely observed. Rates are computed on episodes where \textbf{both} conditions returned a parseable answer, since the estimand is paired; unpaired rates differ by at most 0.2 pp. Adherence collapses in parallel --- 94.4\% $\rightarrow$ \textbf{3.4\%} where the agents were wrong, 95.9\% $\rightarrow$ 58.4\% where they were right (GPT chains). The asymmetry is itself informative: without the field the auditor still lands on an agent-proposed link 58.4\% of the time when that set contains the truth, and 3.4\% when it does not. It is not avoiding the agents' hypotheses; it is no longer confined to them. Overall accuracy rises 26.0\% $\rightarrow$ 35.5\% for GPT chains (MI 0.065 $\rightarrow$ 0.243) and is unchanged for Claude (63.2\% $\rightarrow$ 63.5\%).

\textbf{What this establishes.} The conclusion-bearing field \textbf{causally contributes} to the failure, and reduced information alone cannot explain the gap. It does \textbf{not} establish that the field is the sole cause: recovery reaches 45.2\% against 60.3\% from raw documentation on the same stratum. Nor do we claim explicit evidence for the true origin was present in every report --- the registered instrument for that was never validated (§3.4).

\textbf{Re-collected interleaved.} The two conditions above were issued two days apart, leaving a change in the served model or API as an alternative explanation. We repeated the comparison with both shuffled together under a recorded seed in a \textbf{single execution} (40,278 calls, 4.1\% excluded). The estimates reproduce: \textbf{+41.3} [+35.7, +47.0] and \textbf{+29.0} [+24.7, +33.2] where the suggestion was wrong, -15.7 and -9.2 where it was correct --- each within 2.3 pp of its separate-run figure, three within 1.0 pp. Run-date drift does not explain the effect.

\textbf{Not driven by the omission family, nor by the link-1 accusation.} By defect family, deleting the field helps \textit{more} on inconsistencies than on omissions on both chains: +47.3 against +39.5 pp for GPT-4.1 Mini, +30.7 against +26.4 for Claude. The gain is positive at all five injection positions, every interval excluding zero, and the earliest link --- the one the chains accuse near-constantly --- shows the \textbf{smallest} effect (Appendix J).

\textbf{The trade-off is the finding, not a caveat.} Net benefit depends on the reliability of upstream attribution, which can be estimated offline with the metric in §4.2. We do not recommend removing the field unconditionally.

\textbf{Two controls.} An intervention could raise accuracy by inducing abstention; the abstention rate \textbf{falls}, 34.8\% $\rightarrow$ 26.2\%. The free-text field may itself name the true origin. Excluding every episode whose reports name the true step or the responsible company removes 4 of 5,864 episodes on GPT chains, leaving the gain at +41.2; on Claude chains it removes 31.5\% and the gain falls to \textbf{+20.1} [+16.0, +24.3], still excluding zero. The recovery is not an artifact of residual leakage. The same control on the correct-suggestion stratum is a subgroup analysis on a different estimand (Appendix F).

\subsection{Three replications}
Three replications. The two pre-specified ones were each fixed in a protocol frozen before its own collection; neither belongs to the original pre-registration, and we describe neither as confirmatory.

\textbf{A second auditor.} Both conditions repeated with \texttt{gemini-2.5-flash-lite} on a strict intersection: the gain replicates on both chains, +38.3 [+32.4, +44.3] and +25.9 [+21.5, +30.6] against Qwen's +40.7 and +30.0 on the same episodes, and the cost is directionally consistent in all four cells and significant in three (Appendix I).

\textbf{Frontier auditors.} The auditors above are mid-tier; we repeated both conditions with \texttt{gpt-5} and \texttt{claude-sonnet-5} on 200 episodes per chain $\times$ suggestion-correctness cell.

\textbf{A second domain.} Everything else rests on one environment. We repeated the §4.3 manipulation in a semantically unrelated one --- a six-link \textbf{multi-vendor software delivery toolchain} (source intake, build, test, static analysis, security scanning, release), each link owned by a different vendor so the organisational boundary is preserved. 150 blocks $\times$ 12 cells $\times$ 2 twins $\times$ 6 links = 21,600 requests per chain model, the cell structure of §3.3, prompts built by the Phase 2 path and both conditions by the §4.3 builders (Appendix H).

Where the suggestion was \textbf{wrong}:

\begin{center}
\footnotesize
\setlength{\tabcolsep}{3.0pt}
\begin{tabular}{p{0.179\linewidth}p{0.179\linewidth}p{0.067\linewidth}p{0.112\linewidth}p{0.078\linewidth}p{0.101\linewidth}p{0.157\linewidth}}
\toprule
\textbf{replication} & \textbf{chain} & \textbf{n} & \textbf{with field} & \textbf{without} & \textbf{$\Delta$} & \textbf{95\% CI} \\
\midrule
GPT-5 auditor & GPT-4.1 Mini & 199 & 5.0\% & 28.1\% & \textbf{+23.1} & [+16.4, +30.0] \\
GPT-5 auditor & Claude Haiku 4.5 & 200 & 16.0\% & 24.5\% & \textbf{+8.5} & [+3.7, +13.4] \\
Sonnet 5 auditor & GPT-4.1 Mini & 200 & 13.5\% & 52.5\% & \textbf{+39.0} & [+30.6, +47.2] \\
Sonnet 5 auditor & Claude Haiku 4.5 & 200 & 46.0\% & 76.5\% & \textbf{+30.5} & [+22.6, +38.1] \\
software domain & GPT-4.1 Mini & 514 & 6.8\% & 54.5\% & \textbf{+47.7} & [+39.8, +55.2] \\
software domain & Claude Haiku 4.5 & 113 & 11.5\% & 72.6\% & \textbf{+61.1} & [+41.7, +79.7] \\
\bottomrule
\end{tabular}
\end{center}

\textbf{The harm replicates in six conditions out of six, +8.5 to +61.1 pp, every interval excluding zero.} The frontier reading rule was fixed on accuracy \textit{with} the field --- below 25\% the effect generalises, 25--50\% partial, above 50\% capability-dependent --- and three of four cells fall below 25\%, one is partial at 46.0\%, none exceeds 50\%: capability attenuates the effect in one cell and eliminates it in none. The domain threshold was +15 pp with an interval excluding zero; both readings clear it widely, and the effect is \textbf{larger} than in the original domain.

\textbf{The cost does not replicate.} Across frontier auditors it is heterogeneous --- -30.0 [-38.6, -22.0] and -17.5 [-29.0, -5.4] on GPT chains, -4.5 [-9.5, \textbf{+1.0}] and -1.5 [-7.6, \textbf{+4.4}] on Claude chains, two intervals excluding zero and two not. In the software domain it is absent on both chains: -1.7 [-13.8, \textbf{+9.9}] and +4.0 [-4.0, \textbf{+11.2}]. Without the field the auditor reaches 54.5\% and 72.6\% there against 45.2\% and 35.4\% in domain 1 --- that record carries enough independent signal to locate the origin unaided. \textbf{The trade-off is domain-dependent.} Notably \texttt{gpt-5} recovers \textit{less} than the mid-tier auditor once the field is gone (28.1\% against 45.2\%): greater capability did not produce greater independence.

\subsection{Weak independence (exploratory)}
Under a \textbf{neutral} auditor prompt stating that a defect may or may not exist and that reporting none is correct, run on clean episodes with defective ones batched alongside. Sampling is stratified: all clean episodes with an upstream false alarm (Claude chains produce only 281) plus random samples of the other strata; rates are within stratum, never pooled.

\begin{center}
\footnotesize
\setlength{\tabcolsep}{3.0pt}
\begin{tabular}{p{0.163\linewidth}p{0.265\linewidth}p{0.068\linewidth}p{0.068\linewidth}p{0.153\linewidth}p{0.173\linewidth}}
\toprule
\textbf{chain} & \textbf{stratum} & \textbf{truth} & \textbf{n} & \textbf{claims a defect} & \textbf{\textbf{names a party}} \\
\midrule
GPT-4.1 Mini & defective & defect & 4,000 & 42.1\% & 40.5\% \\
GPT-4.1 Mini & clean, agent false-alarmed & clean & 2,868 & 34.6\% & \textbf{34.4\%} \\
Claude Haiku 4.5 & defective & defect & 3,999 & 40.4\% & 40.4\% \\
Claude Haiku 4.5 & clean, agent false-alarmed & clean & 281 & 74.7\% & \textbf{62.6\%} \\
\textit{both chains} & clean, all agents silent & clean & 7,996 & \textbf{0.0\%} & \textbf{0.0\%} \\
\bottomrule
\end{tabular}
\end{center}

Asserting a defect and naming a party are distinct acts, not conflated. The auditor \textbf{originates almost nothing}: zero allegations across 7,996 clean episodes where every agent stayed silent. It \textbf{filters upstream error poorly}: where an agent false-alarmed it names an innocent party in 34.4\% and 62.6\% of episodes, and for Claude chains its assertion rate there (74.7\%, CI [69.6, 79.8]) exceeds its rate on defective episodes (40.4\%). This is relaying more than checking --- precisely that, not that it filters nothing.

\textbf{A prompt artifact, disclosed.} An earlier run reused the §3.4 auditor prompt, whose second line asserts a defect was introduced --- false of a clean episode. Under it the auditor endorsed the false premise on 62.3\% and 83.3\% of silent clean episodes; that measures compliance, not fabrication, and no result is reported from it. It does yield a format effect: on defective episodes accuracy is 38.2\% under the neutral prompt versus 26.0\% under the asserting one for GPT chains, unchanged for Claude. Asserting the conclusion made the auditor worse at reaching it.

\section{Discussion}
\textbf{What binds.} The limiting factor here is not the auditor's reasoning capability alone --- the same auditor reaches 60.3\% from raw documentation and 4.1\% from reports on the same episodes --- nor simply the volume of information, since deleting one clause recovers 41.2 points. A substantial part of what binds is the \textbf{interface}: the artifact crossing the accountability boundary carries an upstream conclusion alongside its observations, and supplying one while offering little with which to displace it makes adoption likely. No more strongly: recovery stops at 45.2\% against 60.3\%, so the format is one binding factor among others.

\textbf{Why the asymmetry between chains supports this.} Claude chains lose little through the reporting layer (0.809 vs 1.057 bits) while GPT chains lose almost everything (0.065 vs 1.073). GPT-4.1 Mini's omission attributions carry 0.046 of the $\approx$2.3 bits available and name link 1 in 95\% of episodes, a near-constant accusation; Claude's carry 0.715. The auditor inherits whatever hypothesis the chain supplies, so the cost tracks its quality (Appendix B) --- which is why removing the field helps GPT chains and leaves Claude unchanged.

\textbf{A decision rule, not a recommendation.} The intervention is not an improvement to apply by default; it costs 14.7 and 9.0 points where the upstream suggestion was correct. That cost is heterogeneous across auditors (§4.4) and \textbf{absent in the second domain} (§4.4), where the record carries enough independent signal that removing the conclusion takes nothing away. The conditional statement is: \textit{carrying upstream conclusions into an accountability layer supplies useful signal when those conclusions are reliable and misleading anchors when they are not.} The net effect is $\Delta$ = P(wrong)$\cdot$$\Delta$$\mid$wrong + P(correct)$\cdot$$\Delta$$\mid$correct, so prevalence alone does not fix the sign: the two conditional magnitudes are auditor- and domain-dependent (§4.4). What makes this actionable is that upstream reliability \textbf{can be estimated offline} on representative validation data, with the instrument used here, provided the deployment distribution is sufficiently stable. A safer variant than deletion: separate observation from conclusion fields and make the conclusion suppressible at the audit boundary --- available to consumers who benefit, withholdable from the component whose independence it compromises.

\textbf{What it does not establish.} Not that LLM auditors reason poorly --- the same model reaches 60.3\% on raw documentation --- nor that the field is the sole cause, nor that explicit evidence for the true origin was present in every report; and nothing about human organisational behaviour.

\section{Limitations}
\textbf{Two domains, two report schemas} (§4.4) --- the harm replicates in both, the cost in only one, so the trade-off is domain-dependent rather than general. Both environments are synthetic; neither is a deployed pipeline, and each manipulates \textbf{one line}. \textbf{Two chain models}, bounded by the detection screen (§3.8), and \textbf{four auditors, two mid-tier and two frontier} --- this argues against a single-model pathology without establishing invariance. \textbf{A registered specification that did not converge} (§3.6); \textbf{a registered interface that was not implementable} (§3.4); \textbf{a registered analysis not conducted}; \textbf{corpus validation with a single rater}; \textbf{phases specified but not collected} --- the placebo and visibility checks, and a mitigation study void for targeting an effect that does not amplify with chain length. \textbf{The allocation result (Appendix D) is exploratory and partly wording-dependent}, and our own design notes warned that a logical-OR outcome can mask per-unit effects --- it did.

\section{Conclusion}
Our pre-registered hypothesis is not supported on either model. The accountability layer nevertheless fails: an auditor reading the agents' filed reports adopts their proposed origin, and deleting the single clause that carries it recovers 41.2 points where that proposal was wrong, at a cost of 14.7 where it was right. The harm is confined neither to mid-tier auditors nor to one environment; the cost appears in one domain and not the other. The analysis that revealed this was pre-specified as descriptive; the manipulation isolating the cause was exploratory. The design implication: \textit{an accountability layer needs evidence sufficiently independent of the conclusions it verifies; unreliable upstream conclusions convert an error into an authoritative one.}

\section{Reproducibility statement}
The corpus is a pure function of a recorded seed and a generator whose hash is stored with the plan metadata, so it is regenerable without redistribution. Prompt text was frozen before collection and its digest recorded. The pre-registration is timestamped and public prior to confirmatory data collection (DOI 10.17605/OSF.IO/GBR3V). Code and data are available at https://github.com/Polpii/FaultLine. The supplement releases all analysis code, every plan metadata file and generator seed, the frozen protocols and their hashes for the two pre-specified replications of §4.4, and the per-episode derived data behind every reported number. Request plans are deterministically regenerable from the released metadata and seeds. The raw model responses are 5.6 GB and are archived on publication rather than attached here; every figure and table is reproducible from the released derived data. Batch-path prompts were verified byte-identical to the live execution path. Exact model version strings, temperatures, token budgets and structured-output settings are tabulated in Appendix K.

\section{Ethics statement}
This is a simulation using synthetic scenarios; no real company, individual or incident is represented, and no personal data is processed. The dual-use tension is direct: understanding how fault attribution degrades across organisational boundaries could help an actor obscure responsibility as readily as help a defender establish it. We judge publication net-positive because the failure mode is a property of interface design that operators can inspect and correct, and the corresponding defence --- measuring upstream attribution reliability before deciding what the audit interface carries --- is described in enough detail to act on.

\section{Statement on AI assistance}
Generative AI tools were used to assist with language editing, improve clarity and readability, and provide occasional support with code development and literature exploration. All AI-assisted outputs were reviewed and verified by the authors. The authors take full responsibility for the content, analyses, and conclusions presented in this work.

\bibliographystyle{iclr2027_conference}
\bibliography{references}

\appendix
\section{Human corpus validation}
A blind stratified sample of 40 cases was rated against criteria fixed before any answer existed: status accuracy $\geq$ 0.90, unambiguous $\geq$ 0.90, unintended defects $\leq$ 0.05. Difficulty was recorded but explicitly demoted to descriptive before any answer existed, on the reasoning that a rater with unlimited time and a calculator will judge almost everything easy, which says nothing about whether a corpus can carry an effect for models.

\begin{center}
\footnotesize
\setlength{\tabcolsep}{3.0pt}
\begin{tabular}{p{0.331\linewidth}p{0.147\linewidth}p{0.294\linewidth}p{0.147\linewidth}}
\toprule
\textbf{criterion} & \textbf{result} & \textbf{gate} & \textbf{outcome} \\
\midrule
status accuracy & \textbf{100\%} & $\geq$ 0.90 & pass \\
unintended defects & \textbf{0\%} & $\leq$ 0.05 & pass \\
unambiguous & 70\% & $\geq$ 0.90 & \textbf{fail} \\
difficulty band & --- & descriptive only & --- \\
\bottomrule
\end{tabular}
\end{center}

All six failures on the third criterion fell in the omission family. Inspection showed the question conflated two judgements --- whether the stated fact is unambiguous, and whether its materiality is arguable --- so a follow-up on 20 defective cases drawn from the answer key separated them, with the criterion again fixed before answers existed.

\begin{center}
\footnotesize
\setlength{\tabcolsep}{3.0pt}
\begin{tabular}{p{0.336\linewidth}p{0.321\linewidth}p{0.277\linewidth}}
\toprule
\textbf{follow-up criterion} & \textbf{result} & \textbf{outcome} \\
\midrule
stated fact unambiguous & \textbf{100\%} & pass (gate 0.90) \\
materiality arguable & 50\%, \textbf{all omissions} & reported, not gated \\
\bottomrule
\end{tabular}
\end{center}

\textbf{Reading.} The corpus contains real, correctly-labelled defects with no unintended ones. What is arguable, for half of omission cases, is whether the missing record \textit{matters} --- and that is the same family on which one chain model's attribution collapses to 0.046 bits (Appendix B). We report the failed gate and its diagnosis rather than the summary ``validation passed''.

\textbf{Limitation.} A single rater. Inter-annotator agreement is therefore unavailable, and no claim in the paper rests on the difficulty judgement.

\section{Two rejected attribution mechanisms}
Reported because rejected mechanisms are what make the surviving description credible.

\subsection{Self-exculpation}
\textbf{Hypothesis.} Agents are worse at attributing a fault to their own company than to another's.

Raw figures appeared to support it strongly --- accuracy when the defect originated in the agent's own segment versus another's:

\begin{center}
\footnotesize
\setlength{\tabcolsep}{3.0pt}
\begin{tabular}{p{0.275\linewidth}p{0.302\linewidth}p{0.357\linewidth}}
\toprule
\textbf{k} & \textbf{own company} & \textbf{other company} \\
\midrule
3 (GPT) & 13.4\% & 44.3\% \\
6 (GPT) & 6.1\% & 37.2\% \\
3 (Claude) & 42.5\% & 59.6\% \\
6 (Claude) & 18.5\% & 63.8\% \\
\bottomrule
\end{tabular}
\end{center}

\textbf{Why it is an artifact.} ``Own company'' is confounded with distance from the injection point: at k=6 it means the agent \textit{at} the injection link, which sees the defect arrive in its incoming material and naturally attributes it upstream. Controlling for distance d = L - I, the effect does not survive. The decisive control is k=1, where \textbf{no company boundary exists at all}: accuracy at d=1 already ranges from 5.4\% to 97.7\% purely by injection position, and matched positions give near-identical values at k=1 and k=3 (97.7/97.6, 5.4/1.6, 16.4/17.3). Organisational structure explains nothing once position is held fixed.

\subsection{Semantic-role attribution}
\textbf{Hypothesis.} Agents blame the role whose responsibility the defect's wording implicates, rather than the position where it entered.

The omission defect text names a document type drawn at random, independently of injection position --- an unbalanced natural crossing of exactly the two factors. If the hypothesis held, blame would follow the document type.

\begin{center}
\footnotesize
\setlength{\tabcolsep}{3.0pt}
\begin{tabular}{p{0.542\linewidth}p{0.408\linewidth}}
\toprule
\textbf{document type named} & \textbf{GPT blames link 1} \\
\midrule
quality control sign-off & 97\% \\
dimensional inspection record & 98\% \\
material certificate of conformity & 96\% \\
final release check & 88\% \\
\bottomrule
\end{tabular}
\end{center}

It does not. Blame tracks neither the true position nor the semantic role; it is close to constant.

\subsection{Chain-agent attribution}
GPT-4.1 Mini's omission attributions carry 0.046 of the $\approx$2.3 bits available and name link 1 in 95\% of episodes, a near-constant accusation; Claude's carry 0.715 bits. This supplies the auditor's hypothesis space and explains the §4.2 asymmetry: Claude chains lose little because their suggestions are informative to begin with. Per-family figures and two rejected mechanisms (self-exculpation, semantic-role attribution) are in Appendix B. Abstention was available and unused: \texttt{origin\_estimate} is left null in 0.2--0.3\% of escalations.

\section{Model screening}
Every candidate was run on \textbf{real chain prompts} with the real schema-forced output, not a toy prompt.

\textbf{Structured output.} Seven of eight candidates held the forced schema on 3/3 strict-mode trials with zero truncation. \texttt{deepseek-v4-flash} passed 2/3 --- one truncated response and one empty completion --- and was excluded rather than special-cased.

\textbf{Detection}, per-link escalation on k=6 defective episodes, 300 calls each:

\begin{center}
\footnotesize
\setlength{\tabcolsep}{3.0pt}
\begin{tabular}{p{0.632\linewidth}p{0.173\linewidth}p{0.130\linewidth}}
\toprule
\textbf{model} & \textbf{rate} & \textbf{usable} \\
\midrule
gpt-4.1-mini \textit{(reference, used as chain model)} & 17.1\% & --- \\
claude-haiku-4.5 \textit{(reference, used as chain model)} & 12.5\% & --- \\
mistral-small-3.2-24b & 12.3\% & yes \\
gemini-2.5-flash-lite & 9.3\% & yes \\
gemma-3-27b & 4.7\% & no \\
qwen3-30b-a3b & 3.7\% & no \\
mistral-nemo & \textbf{0.0\%} & no \\
llama-3.3-70b & \textbf{0.0\%} & no \\
gpt-4.1-nano & \textbf{0.0\%} & no \\
\bottomrule
\end{tabular}
\end{center}

Three models returned \texttt{escalate: false} on all 300 calls. This was verified as model behaviour rather than a pipeline fault: all seven produced complete records with coherent free text (``No irregularities found in the provided documentation and logs'').

\textbf{Interpretation.} The corpus difficulty was calibrated against the two chain models used here and does not transfer downward. A study of attribution cannot be run on chains that produce almost no attributions, so these models are excluded --- reported here as a bound on where the result can be measured rather than folded silently into model selection.

\section{The allocation effect itself}
Moved here from the main text: the registered contrast is null, and this is the descriptive gap it did not test. It is exploratory and partly wording-dependent, and the wording breakdown follows immediately below.

The registered contrast tested whether the allocation gap widens with \textit{k}. It does not. The gap itself, never registered as a contrast, is substantial and flat across \textit{k}. J-gap (individual minus collective), block-clustered bootstrap: responsibility +3.8\% [+1.9, +5.8] and +5.9\% [+4.0, +7.5]; documentation -4.6\% [-6.1, -2.8] and -4.7\% [-5.8, -3.5], for GPT and Claude chains. Flatness across \textit{k} (GPT responsibility: +3.8/+3.6/+4.0 at k=1/3/6) is precisely why a contrast testing amplification is null: the registered hypothesis assumed the wrong functional form. \textbf{This is not a confirmation of the original hypothesis}; the two claims are kept separate.

\textbf{The effect is wording-dependent for one model}: robust across all three variants in the documentation domain and for Claude, but for GPT-4.1 Mini not a discrimination effect outside the most emphatic phrasing (Appendix E). A confirmatory registration is prepared and unrun.

\section{Wording-variant breakdown of the allocation effect}
The allocation manipulation is one paragraph of framing text with three lexical variants rotated across blocks. The leave-one-wording-out check required by our own design notes is reported here in full, including where it fails.

J-gap (individual minus collective), block-clustered bootstrap:

\begin{center}
\footnotesize
\setlength{\tabcolsep}{3.0pt}
\begin{tabular}{p{0.256\linewidth}p{0.221\linewidth}p{0.175\linewidth}p{0.268\linewidth}}
\toprule
\textbf{} & \textbf{v0} & \textbf{v1} & \textbf{v2} \\
\midrule
GPT, responsibility & \textbf{+9.3\%} (excl. 0) & +0.8\% (incl. 0) & +1.4\% (incl. 0) \\
Claude, responsibility & \textbf{+9.6\%} & \textbf{+5.2\%} & \textbf{+2.8\%} (all excl. 0) \\
GPT, documentation & \textbf{-3.6\%} & \textbf{-5.5\%} & \textbf{-4.5\%} (all excl. 0) \\
Claude, documentation & \textbf{-4.7\%} & \textbf{-3.0\%} & \textbf{-6.2\%} (all excl. 0) \\
\bottomrule
\end{tabular}
\end{center}

\textbf{The documentation-domain effect is robust} across all three wordings on both models. \textbf{The responsibility-domain effect is not}, for GPT-4.1 Mini: it is present on raw escalation (v1 +5.2 pp, v2 +4.1 pp on detection) but the matched rise in false alarms means it is not a discrimination effect outside v0.

The variants differ in emphasis rather than content. v0 stacks three emphasis clauses --- \textit{``personally \textbf{and} individually responsible\ldots \textbf{regardless of} how many other companies\ldots \textbf{yours alone and does not depend on what anyone else does}''} --- where v1 and v2 carry one. The effect tracks emphasis intensity, which is an interpretable pattern but one fitted post hoc to three points, and we do not present it as a dose-response finding.

The registered contrast on which the confirmatory claim would rest survives excluding v0: +6.12 pp (z = 3.65) for GPT chains and +8.62 pp (z = 7.68) for Claude, because the documentation component carries it.

\section{Free-text leakage control on the correct-suggestion stratum}
The control reported in §4.3 covers the wrong-suggestion stratum, where it answers the objection that the recovery might come from the free text rather than from the deleted field. The same control on the \textbf{correct}-suggestion stratum is reported here rather than in the main text, because it is a subgroup analysis on a different estimand and does not replace the full-sample full-sample result of §4.3.

Episodes are excluded when any escalating agent's free text names the true injection step or the responsible company. Paired block-clustered bootstrap, 2,000 resamples.

\begin{center}
\footnotesize
\setlength{\tabcolsep}{3.0pt}
\begin{tabular}{p{0.182\linewidth}p{0.204\linewidth}p{0.068\linewidth}p{0.114\linewidth}p{0.079\linewidth}p{0.068\linewidth}p{0.159\linewidth}}
\toprule
\textbf{chain} & \textbf{population} & \textbf{n} & \textbf{with field} & \textbf{without} & \textbf{$\Delta$} & \textbf{95\% CI} \\
\midrule
GPT-4.1 Mini & all paired (§4.3) & 3,988 & 70.5\% & 55.7\% & -14.7 & [-25.5, -4.9] \\
GPT-4.1 Mini & no leakage (95.4\%) & 3,806 & 69.5\% & 54.2\% & -15.3 & [-25.3, -5.0] \\
Claude Haiku 4.5 & all paired (§4.3) & 6,439 & 80.6\% & 71.6\% & -9.0 & [-12.6, -5.6] \\
Claude Haiku 4.5 & no leakage (40.3\%) & 2,595 & 79.4\% & 54.3\% & -25.2 & [-31.5, -18.8] \\
\bottomrule
\end{tabular}
\end{center}

On GPT chains the subset is 95.4\% of the sample and the estimate is unchanged. On Claude chains it is \textbf{40.3\%} of the sample, and the larger cost there is a statement about that subgroup, not a correction of the -9.0 full-sample figure. The subgroup is defined by the agents' own output, so conditioning on it selects a subpopulation rather than removing a bias --- the same caution that applies to the stratification in §4.2.

The direction is nevertheless mechanically interpretable: where the free text already carries the origin, the field is redundant and deleting it costs little; where the free text is silent, the field is the only carrier of a conclusion, and deleting it costs a great deal. This is what the causal claim predicts.

\section{Alternative variance specifications for the pre-registered contrast}
The pre-registered working correlation (exchangeable) did not converge on either chain model, and the reported analysis falls back to independence with the same Mancl-DeRouen bias-corrected sandwich. The primary contrast under four specifications, on the same data and the same estimand:

\begin{center}
\footnotesize
\setlength{\tabcolsep}{3.0pt}
\begin{tabular}{p{0.335\linewidth}p{0.384\linewidth}p{0.216\linewidth}}
\toprule
\textbf{specification} & \textbf{GPT-4.1 Mini} & \textbf{Claude Haiku 4.5} \\
\midrule
GEE exchangeable (pre-registered) & +0.22 [-3.90, +4.34] --- did not converge & did not converge \\
GEE independence (reported) & +1.08 [-3.05, +5.22] & -1.29 [-4.11, +1.52] \\
GLMM, random block intercept (VB) & +1.66 [-5.33, +8.66] & -3.83 [-8.55, +0.89] \\
GLMM, random block intercept (MAP) & +0.20 [-30.98, +31.37] & -5.69 [-26.65, +15.27] \\
\bottomrule
\end{tabular}
\end{center}

All four intervals include zero: the null result is not an artifact of the fallback. The estimated between-block standard deviation is 4.1 to 7.8 on the logit scale, which is the likely reason a single exchangeable within-cluster correlation could not be fit. The variational-Bayes intervals assume posterior independence between fixed effects and are approximate; the point estimates and the sign of the conclusion are what these fits establish, not the exact interval widths.

\section{Second domain: corpus construction and saturation screen}
The replication of §4.4 required a corpus in a second domain. This appendix records how it was built and, in particular, what the pre-collection screen rejected, because the rejections are more informative than the survivors.

\textbf{Design.} Six-link multi-vendor software delivery toolchain. Injection at exactly one link 1..5, positions exactly balanced by construction (30 blocks per position), case content a pure function of a recorded seed. Prompts assembled through the same MCP path as §3.2; the frozen prompt digest is identical to the one used for the main collection.

\textbf{Saturation screen.} The protocol fixed the rule before collection: a mechanism whose per-link escalation rate is at or above 90\%, or at or below 5\%, on either chain model is excluded. 50 cases per mechanism, both chain models, presented to the first downstream observer.

\begin{center}
\footnotesize
\setlength{\tabcolsep}{3.0pt}
\begin{tabular}{p{0.318\linewidth}p{0.201\linewidth}p{0.268\linewidth}p{0.134\linewidth}}
\toprule
\textbf{mechanism} & \textbf{GPT-4.1 Mini} & \textbf{Claude Haiku 4.5} & \textbf{outcome} \\
\midrule
artifact\_files & 42.0\% & 24.0\% & kept \\
symbol\_sum & 0.0\% & 0.0\% & excluded \\
dependency\_version & 36.0\% & 100.0\% & excluded \\
coverage\_regression & 2.0\% & 2.0\% & excluded \\
commit\_mismatch & 30.0\% & 90.0\% & excluded \\
missing\_signature & 76.0\% & 86.0\% & kept \\
\bottomrule
\end{tabular}
\end{center}

\textbf{An anchoring assumption that failed.} Two mechanisms were designed to reproduce the supply-chain arithmetic exactly, so that difficulty would be comparable across domains by construction: short by one container, a 1/n shortfall verified across 1,200 generated pairs to fall in the same 3.8-5.6\% band. In the first screen both measured \textbf{0.0\%} per-link escalation on GPT chains. The failure was semantic, not arithmetic --- the first version compared a registered test \textit{plan} against test \textit{execution}, and a gap between the two reads as routine in software, so there is nothing to flag. Rebuilt around quantities that must reconcile exactly in a software artifact, the same arithmetic moved from 0.0\% to 42.0\%. \textbf{The shortfall ratio transfers across domains; detectability does not.}

\textbf{A borderline case, decided against ourselves.} \texttt{commit\_mismatch} measured 88.0\% in the first screen and 90.0\% in the second on Claude chains, straddling the exclusion threshold; at n = 50 the standard error is roughly 4 pp, so the two are not distinguishable. We applied the rule as written and excluded it, rather than re-screening at larger n until the answer changed.

\textbf{Corpus actually built.} Two mechanisms survive, one per defect family: \texttt{artifact\_files} (inconsistency) and \texttt{missing\_signature} (omission). This is the same two-family structure as domain 1, which also retained two families after excluding one for saturating at 100\% per link. Measured chain behaviour on the collected corpus: detection 67.2\% and 52.7\%, false-alarm 21.3\% and 14.1\%, giving J = +45.9 and +38.7 points; per-link escalation 28.4-28.5\% against the domain-1 references of 17.1\% and 12.5\%, so domain 2 is somewhat more detectable --- a difference reported, not corrected for.

\textbf{Auditor exclusions.} 440 of 7,200 auditor responses (6.1\%) failed to parse or returned an origin outside 1..6 and are excluded, matching the rule used throughout.

\section{Second-auditor replication, full table}
Both conditions of §4.3 repeated with \texttt{gemini-2.5-flash-lite}, on the strict intersection where both auditors returned a parseable answer in both conditions. Summarised in §4.4; reproduced here in full.

\begin{center}
\footnotesize
\setlength{\tabcolsep}{3.0pt}
\begin{tabular}{p{0.213\linewidth}p{0.133\linewidth}p{0.280\linewidth}p{0.293\linewidth}}
\toprule
\textbf{chain} & \textbf{suggestion} & \textbf{Qwen3-30B: $\Delta$ [95\% CI]} & \textbf{Gemini: $\Delta$ [95\% CI]} \\
\midrule
GPT-4.1 Mini & wrong & +40.7 [+34.9, +46.7] & +38.3 [+32.4, +44.3] \\
GPT-4.1 Mini & correct & -15.2 [-26.6, -5.3] & -7.1 [-18.3, \textbf{+3.8}] \\
Claude Haiku 4.5 & wrong & +30.0 [+24.2, +35.4] & +25.9 [+21.5, +30.6] \\
Claude Haiku 4.5 & correct & -7.2 [-11.5, -3.2] & -5.9 [-9.1, -2.6] \\
\bottomrule
\end{tabular}
\end{center}

Seven of eight intervals exclude zero.

\section{Defect family, injection position, and the link-1 accusation}
Reported in §4.3 in summary. Both breakdowns are \textbf{exploratory analyses of already collected data, performed after review}, on the wrong-suggestion stratum of §4.3.

\begin{center}
\footnotesize
\setlength{\tabcolsep}{3.0pt}
\begin{tabular}{p{0.187\linewidth}p{0.152\linewidth}p{0.069\linewidth}p{0.117\linewidth}p{0.082\linewidth}p{0.105\linewidth}p{0.163\linewidth}}
\toprule
\textbf{chain} & \textbf{defect family} & \textbf{n} & \textbf{with field} & \textbf{without} & \textbf{$\Delta$} & \textbf{95\% CI} \\
\midrule
GPT-4.1 Mini & inconsistency & 1,255 & 16.3\% & 63.6\% & \textbf{+47.3} & [+37.8, +56.3] \\
GPT-4.1 Mini & omission & 4,609 & 0.7\% & 40.2\% & +39.5 & [+32.7, +46.3] \\
GPT-4.1 Mini & \textit{both} & 5,864 & 4.1\% & 45.2\% & +41.2 & [+35.3, +46.8] \\
Claude Haiku 4.5 & inconsistency & 127 & 33.9\% & 64.6\% & \textbf{+30.7} & [+19.0, +40.4] \\
Claude Haiku 4.5 & omission & 2,280 & 7.3\% & 33.8\% & +26.4 & [+22.1, +30.5] \\
Claude Haiku 4.5 & \textit{both} & 2,407 & 8.7\% & 35.4\% & +26.7 & [+22.7, +30.6] \\
\bottomrule
\end{tabular}
\end{center}

The concern this answers is specific: GPT-4.1 Mini's omission attributions carry 0.046 bits and name link 1 in 95\% of episodes (Appendix B), so an effect concentrated there would be an artifact of one degenerate behaviour rather than a property of the interface. The measured effect is \textbf{larger on inconsistencies than on omissions} on both chains, so the omission family dilutes the pooled figure rather than producing it.

By injection position, GPT chains:

\begin{center}
\footnotesize
\setlength{\tabcolsep}{3.0pt}
\begin{tabular}{p{0.226\linewidth}p{0.081\linewidth}p{0.162\linewidth}p{0.113\linewidth}p{0.081\linewidth}p{0.226\linewidth}}
\toprule
\textbf{injection link} & \textbf{n} & \textbf{with field} & \textbf{without} & \textbf{$\Delta$} & \textbf{95\% CI} \\
\midrule
1 & 393 & 12.5\% & 21.9\% & +9.4 & [+3.2, +17.4] \\
2 & 1,256 & 1.3\% & 8.8\% & +7.5 & [+3.3, +13.1] \\
3 & 1,367 & 10.6\% & 42.2\% & +31.6 & [+23.2, +41.5] \\
4 & 1,027 & 1.3\% & 52.8\% & +51.5 & [+41.9, +59.1] \\
5 & 1,821 & 0.8\% & 73.5\% & +72.7 & [+65.3, +79.0] \\
\bottomrule
\end{tabular}
\end{center}

Every interval excludes zero. Link 1 --- the position the chains accuse near-constantly --- carries the smallest effect, not the largest. The gradient is driven by the without-field arm (21.9\% rising to 73.5\%) while the with-field arm stays between 0.8\% and 12.5\% at every position: the field suppresses attribution regardless of where the defect sits, and what varies is how recoverable the origin is once the field is gone. That pattern is consistent with recency of the evidence in the accumulated record; we did not manipulate position causally and do not claim it as the explanation.

\section{Model versions and sampling settings}
The Reproducibility statement points here. Every call in the study, with the settings actually used.

\begin{center}
\footnotesize
\setlength{\tabcolsep}{3.0pt}
\begin{tabular}{p{0.216\linewidth}p{0.272\linewidth}p{0.128\linewidth}p{0.136\linewidth}p{0.152\linewidth}}
\toprule
\textbf{role} & \textbf{model string} & \textbf{temperature} & \textbf{max output tokens} & \textbf{structured output} \\
\midrule
chain agents & \texttt{gpt-4.1-mini} & 0.7 & 512 & JSON schema, strict \\
chain agents & \texttt{claude-haiku-4-5-20251001} & 0.7 & 512 & forced tool call \\
auditor, primary (§4.2--4.3) & \texttt{qwen/qwen3-30b-a3b-instruct-2507} & 0.3 & 900 & JSON schema, strict \\
auditor, replication (§4.4) & \texttt{google/gemini-2.5-flash-lite} & 0.3 & 900 & JSON schema, strict \\
auditor, frontier (§4.4) & \texttt{gpt-5} & provider default & 6000 & JSON schema, strict \\
auditor, frontier (§4.4) & \texttt{claude-sonnet-5} & provider default & 6000 & forced tool call \\
\bottomrule
\end{tabular}
\end{center}

Temperature is not passed to the two frontier auditors: \texttt{claude-sonnet-5} rejects the parameter and \texttt{gpt-5} does not accept it on this endpoint. The 6000-token cap replaced an initial 4000 after a pilot measured one response in six truncating at that limit, which bills the call and yields nothing. Information-boundary conditions are \texttt{visibility = local\_only} and \texttt{identity = named} throughout.

\section{Supplementary figures}
Supplementary visualisations of results reported in the main text.

\begin{figure}[t]
\centering
\includegraphics[width=\linewidth]{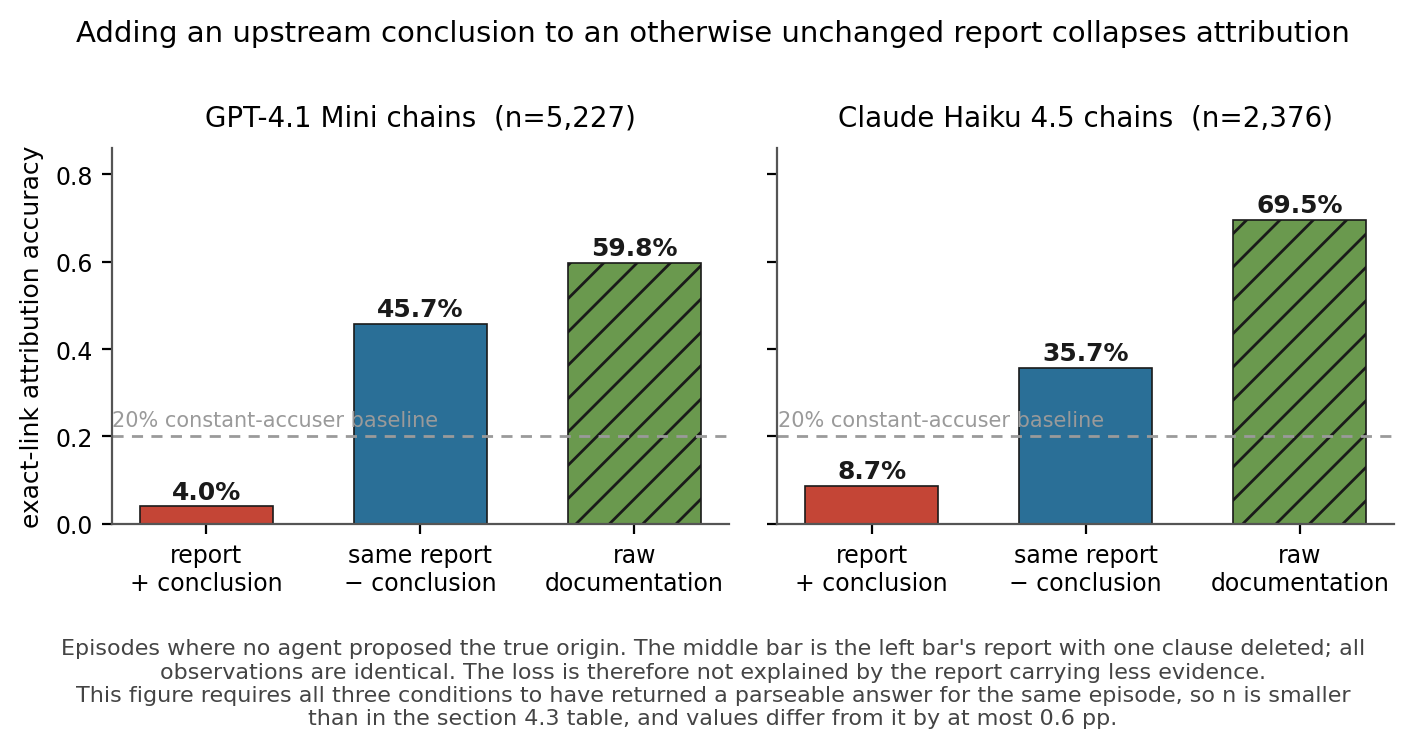}
\caption{Same auditor, same episodes: a report carrying an upstream conclusion, that report with the conclusion deleted, and the raw documentation. Complete-case subset across all three conditions.}
\label{fig:three}
\end{figure}

\begin{figure}[t]
\centering
\includegraphics[width=\linewidth]{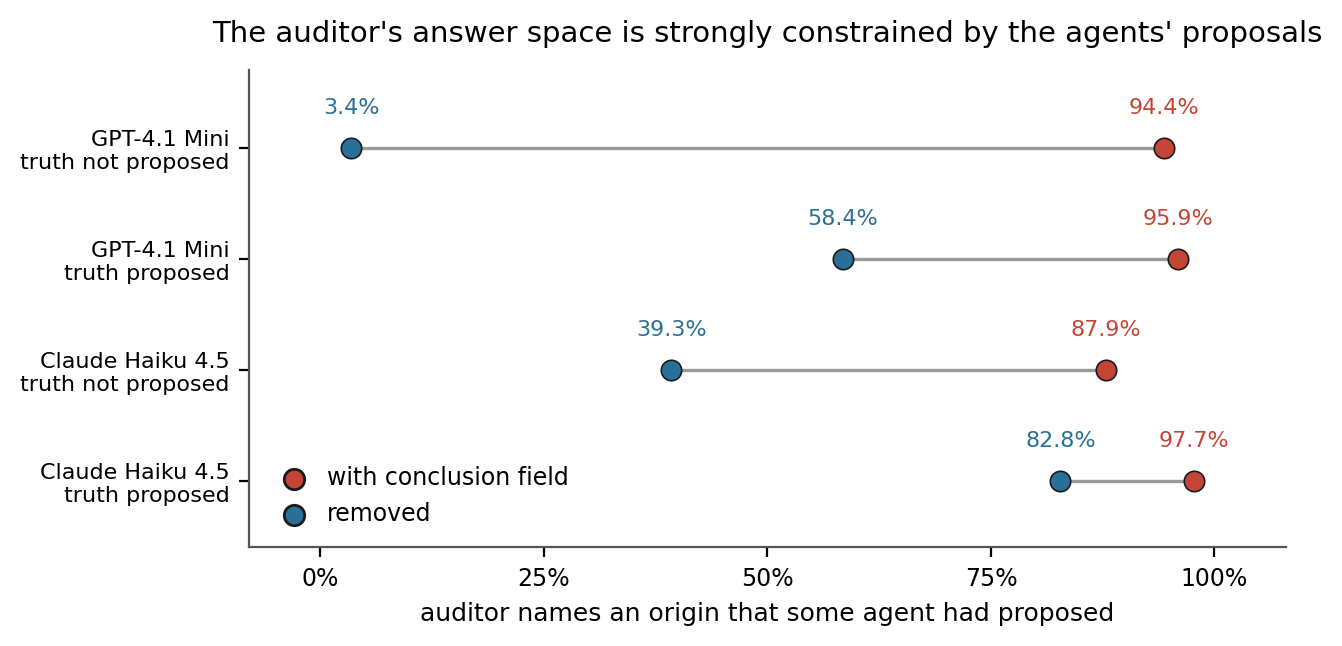}
\caption{Adherence to the agents' proposed origins, by stratum, with and without the conclusion field.}
\label{fig:adh}
\end{figure}

\begin{figure}[t]
\centering
\includegraphics[width=\linewidth]{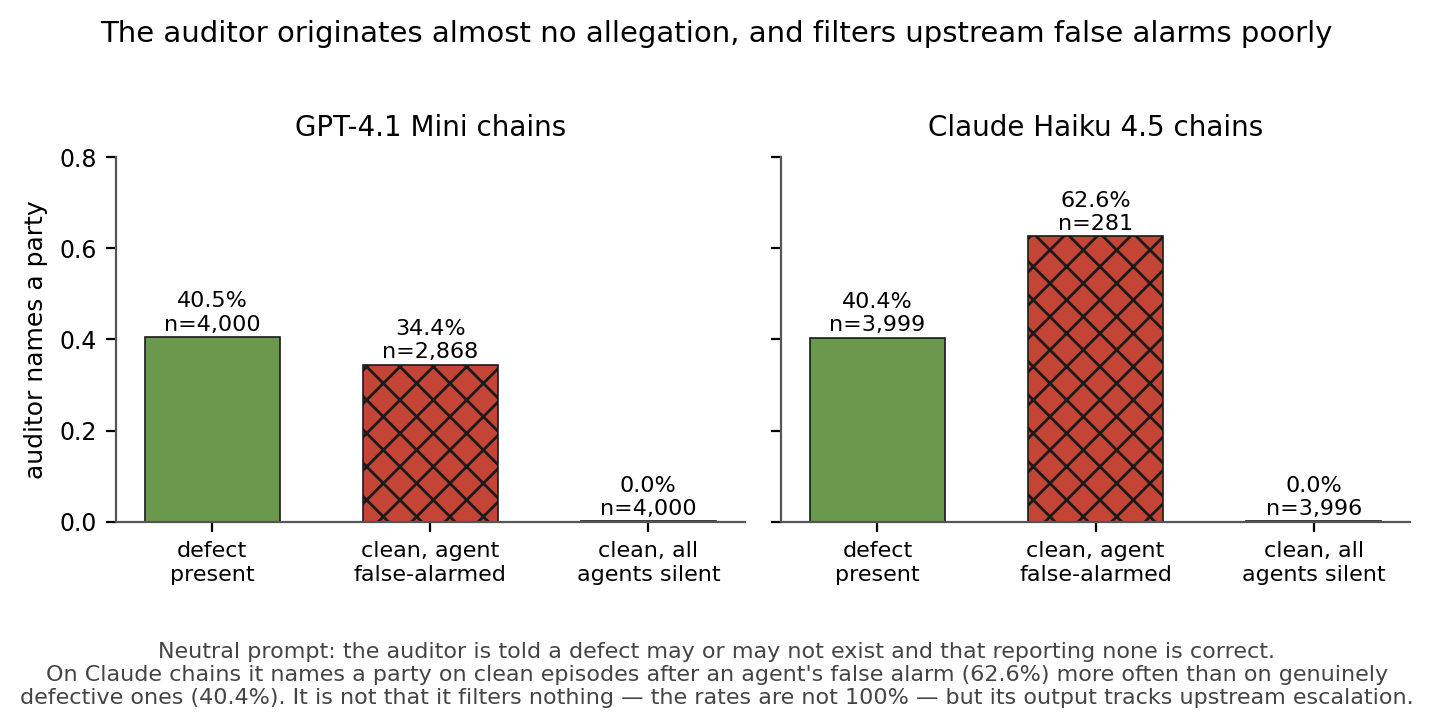}
\caption{Under a neutral prompt the auditor originates almost no allegation without upstream escalation, yet frequently endorses an upstream false alarm on a clean episode.}
\label{fig:weak}
\end{figure}

\begin{figure}[t]
\centering
\includegraphics[width=\linewidth]{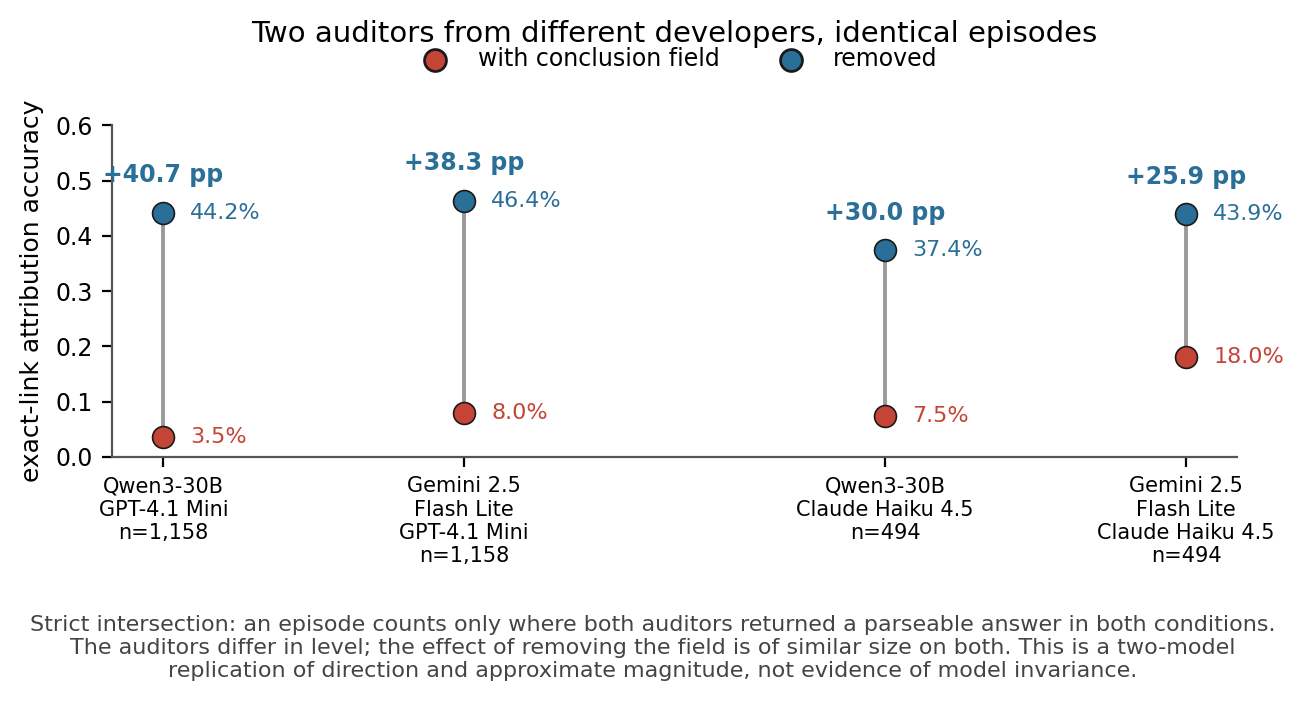}
\caption{Second-auditor replication on the strict intersection of episodes.}
\label{fig:rep}
\end{figure}

\end{document}